\documentclass{icrc2009}

\usepackage{graphicx}   % for including figures
\usepackage{caption}    % for captions
\usepackage[font=footnotesize]{subfig} % subfig.sty for a double column floating figure using two subfigures
\usepackage{fixltx2e}
\usepackage{url}

\newcommand{\shorttitle}[1]%
{\markboth{Proceedings of the 31\MakeLowercase{$^{st}$} ICRC, {\L}\'{o}d\'{z} 2009}{#1} }
\newcommand{\etal}{\MakeLowercase{\textit{et al. }}} % "et al."

\newcommand{\antares}{ANTARES}
\begin{document}
\title{Search for gamma-ray bursts with the \antares\ neutrino telescope}

\author{\IEEEauthorblockN{Mieke Bouwhuis\IEEEauthorrefmark{1} 
			on behalf of the \antares\ collaboration}
\IEEEauthorblockA{\IEEEauthorrefmark{1}National Institute for
Subatomic Physics (Nikhef), Amsterdam, The Netherlands}}

% please write the preseter's name and short title (3-4 words maximum)
%    which will appear at the header of the even pages.
\shorttitle{Bouwhuis \etal GRB search \antares}
\maketitle

\begin{abstract}
Satellites that are capable of detecting gamma-ray bursts can trigger
the \antares\ neutrino telescope via the real-time gamma-ray bursts 
coordinates network.
Thanks to the ``all-data-to-shore" concept that is implemented
in the data acquisition system of \antares, the sensitivity to
neutrinos from gamma-ray bursts is significantly increased when
a gamma-ray burst is detected by these satellites.
The performance of the satellite-triggered data taking is shown,
as well as the resulting gain in detection efficiency.
Different search methods can be applied to the data taken in
coincidence with gamma-ray bursts.
For gamma-ray bursts above the \antares\ horizon, for which a neutrino 
signal is more difficult to find, an analysis method is
applied to detect muons induced by the high-energy gamma rays
from the source. 
  \end{abstract}


\section{Introduction}

Several models predict the production of high-energy neutrinos by
gamma-ray bursts (GRBs)~\cite{grbs}.
The detection of these neutrinos would provide evidence for hadron
acceleration by GRBs. 
Such an observation would lead to a better understanding of the extreme
processes associated with these astrophysical phenomena.
In particular, it would give insight into the creation and
composition of relativistic jets.

One of the goals of the \antares\ neutrino telescope is to detect
high-energy neutrinos from GRBs.
The \antares\ telescope is situated in the Mediterranean Sea
at a depth of about 2500~m.
Neutrinos are detected through the detection of Cherenkov light
induced by the charged lepton that emerges from a neutrino
interaction in the vicinity of the detector.
Measurements are focused mainly on muon-neutrinos, since the muon 
resulting from a neutrino interaction can travel a distance of up to 
several kilometres.
Due to the transparency of the sea water (the absorption length is about
50~m), the faint Cherenkov light can be detected at relatively large
distances from the muon track.
A large volume of sea water is turned into a neutrino
detector by deploying a 3-dimensional array of light sensors in the
water.
The instrumented volume of sea water in the \antares\ telescope 
amounts to about 50 million cubic metres.
The track of the muon can be reconstructed from the measured arrival
times of the Cherenkov photons at the photo-multiplier tubes.
Since the muon and neutrino paths are approximately co-linear at high energies, 
the direction of the neutrino, and thus its origin, can be determined.

The GRB Coordinates Network (GCN)~\cite{gcn} announces the occurrence
of a GRB by distributing real-time alerts.
The data acquisition system of the \antares\ detector is designed such
that it can trigger in real time on these alerts.
This increases the detection efficiency for neutrinos from GRBs
significantly.
The \antares\ detector is currently the only neutrino telescope that
can trigger in real time on GRB alerts.

\section{Data taking with the \antares\ detector}
\label{s:data_taking}
The \antares\ telescope is operated day and night.
During operation, all signals from the photo-multiplier tubes are
digitised, and the raw data (containing the charge and time information of detected
Cherenkov photons) are sent to shore in a continuous data stream.
This is known as the all-data-to-shore concept~\cite{daq}.
Although daylight does not penetrate to the depth of the \antares\
site, a ubiquitous background luminosity is present in the deep-sea
due to the decay of radioactive isotopes (mainly $^{40}$K) and
to bioluminescence.
This background luminosity produces a relatively high count rate of
random signals in the detector (60--150~kHz per photo-multiplier tube).
The total data rate is primarily determined by this background
luminosity, and amounts to about 1 GB/s.
On shore, the continuous data stream is divided over a farm of PCs.
Each of these PCs has a sophisticated filter program running, which
processes the data it receives in real time.
This filter scans the full sky, and finds the correlated photons that
are caused by a muon traversing the detector.
It triggers at a threshold of 10 such photons, which translates to a high
detection efficiency for muons, while preserving a high muon purity
(better than 90\%).
At the average background rate, the total trigger rate is 5--10~Hz
(depending on the trigger conditions).
The data are effectively reduced by a factor of 10$^4$.

\section{Satellite triggered data taking}

The data acquisition system of the \antares\ detector is linked with a socket 
connection to the GCN. 
The GCN network includes the Swift and Fermi satellites, both capable
of detecting GRBs.
When a GRB alert is received from the GCN, the standard data
processing continues (described in section~\ref{s:data_taking}), and in parallel 
to that the satellite triggered data taking is applied: all raw data 
covering a preset period (presently 2~minutes) are saved to disk for each GRB alert. 
There are about 1--2 GRB alerts per day, and half of them correspond to
a real GRB.

The buffering of the data in the data filter processors is used to store
the data up to about one minute before the actual alert.
The amount of data that can be kept in memory depends on the 
background rate in the sea water, the number of data processing PCs,
and the size of the RAM.
These data not only cover the delay between the detection of the GRB
by the satellite and the arrival time of the alert at the
\antares\ site, but also include data collected by the \antares\
detector before the GRB occurred.
These data therefore include a possible early neutrino signal that is
observable before the gamma rays. 

\begin{figure}[t!]
\centering
\includegraphics[width=3.0in]{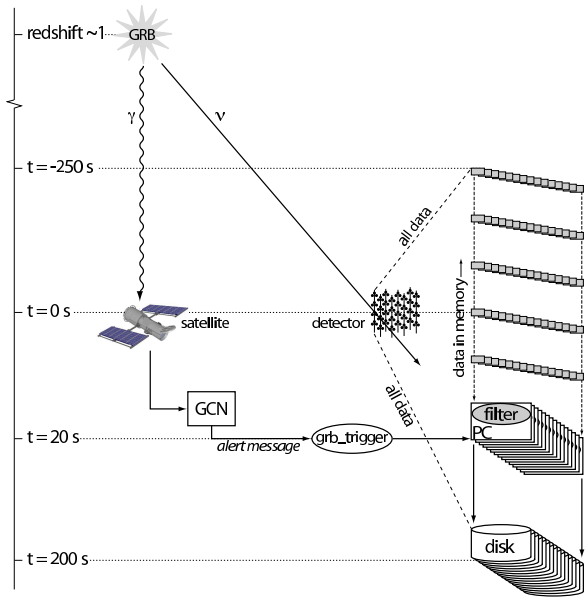}
\caption{Time line of the different events during a GRB. When an alert 
from the GCN is received by \antares\ (grb\_trigger program),
all raw data covering a few minutes are saved to disk, including all
data in memory. Any neutrino signal from the GRB (before, during, and
after the photon detection by the satellite) is stored on disk.}
\label{fig:fig01}
\end{figure}

For each GRB that is detected by a satellite, and announced by
the GCN, all raw data collected by the \antares\ detector in
coincidence with the GRB are available on disk, as shown schematically
in Fig.~\ref{fig:fig01}.
The satellite triggered data taking period per GRB
 corresponds to a few times the typical duration of a GRB.
As a result, any time-correlated neutrino signal from the GRB
---before, during, and after the photon detection by the satellite---
is stored on disk.
Saving all raw data is only possible for transient sources like GRBs.
It can not be done for continuous sources because of the high data 
output rate of the detector.

\section{Satellite triggered data taking performance}

\begin{figure}[t!]
\setlength{\unitlength}{1cm}
\begin{center}
\begin{picture}(8,8.5)
\put(0,0){\scalebox{0.45}{\includegraphics{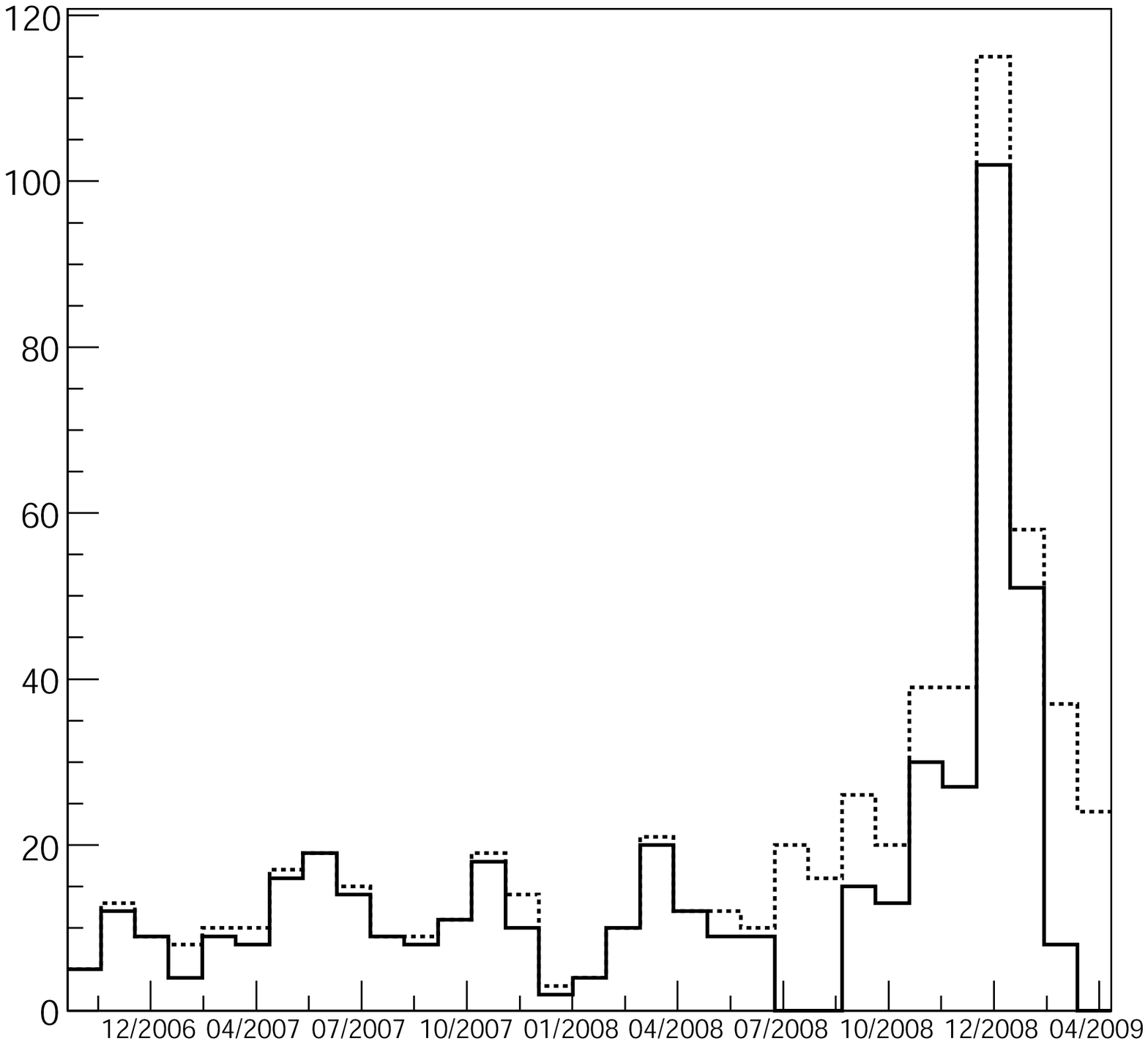}}}%
\put(4.5,0.1){\makebox(0,0)[b]{date (mm/yyyy)}}%
\put(0,4.5){\makebox(0,0)[l]{\rotatebox{90}{number of GRB alerts}}}%
\end{picture}
\end{center}
\caption{The satellite triggered data taking of \antares\
since the implementation of the satellite triggered data taking system.
The dashed line indicates the number of GRB alerts distributed
by the GCN per month. 
The solid line indicates the number of GRB alerts for which
the satellite triggered data taking was applied.
In November~2008, the GCN started the distribution of alerts from the 
Fermi satellite.
The relatively large amount of alerts received in January and 
February~2009 is due to the flaring activity of SGR~1550-5418.} 
\label{fig:fig02}
 \end{figure}

The satellite triggered data taking for all GRB alerts
distributed by the GCN is shown in Fig.~\ref{fig:fig02}.
The satellite triggered data taking system became operational in 
autumn 2006.  
The dashed line shows the number of GRB alerts from the GCN per
month as a function of time, and the solid line shows the
number of satellite triggered data taking sessions that were realised.
Although data taking with \antares\ is in principle continuous, 
inefficiencies can occur, for example, when a GRB alert is 
distributed during a calibration run, or due to power loss.  
As can be seen from Fig.~\ref{fig:fig02}, the typical efficiency of
the satellite triggered data taking is about 90\%. 

\begin{figure}[t!]
\setlength{\unitlength}{1cm}
\begin{center}
\begin{picture}(8,5.5)
\put(0.5,0){\scalebox{0.45}{\includegraphics{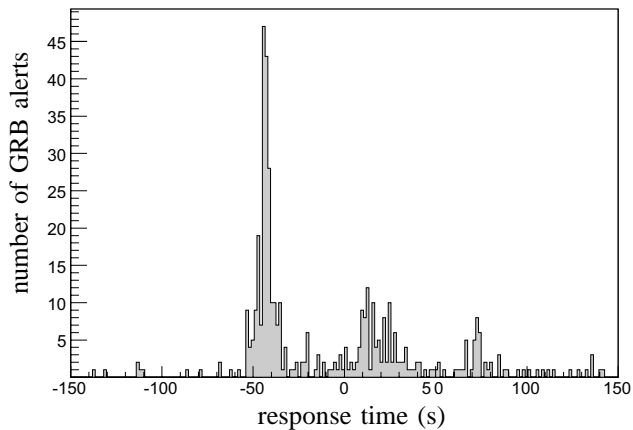}}}%
\put(4.5,-0.4){\makebox(0,0)[b]{response time (s)}}%
\put(0,3){\makebox(0,0)[l]{\rotatebox{90}{number of GRB alerts}}}%
\end{picture}
\end{center}
\caption{Earliest datum in the unfiltered GRB data set relative to the
detection time of the GRB by the satellite. At negative response
times, the unfiltered data set includes data prior to the detection of
the GRB by the satellite. Positive response times indicate a
substantial delay between the detection of the GRB by the satellite,
and the arrival time of the alert at the site.}
\label{fig:fig03}
\end{figure}

The response time of the \antares\ satellite triggered data taking to
the detection of the GRB by the satellite is shown in
Fig.~\ref{fig:fig03}.
The response time is defined as the time difference between the GRB
time, as given in the GCN alert message, and the earliest datum in the
unfiltered data set available on disk. 
This indicates the amount of overlap of the unfiltered data set
with the observation period of the GRB by the satellite (the satellite
triggered data taking lasts for a fixed period of time).
At a response time of 0~seconds, the data in the unfiltered data set
completely cover the period during which the GRB was detected by the
satellite.
For positive response times, the delay between the detection of the GRB
by the satellite and the arrival time of the alert at the
\antares\ site plays a role.
As a result, the unfiltered data saved on disk do not fully cover the
period during which the GRB was detected by the satellite.
For negative response times, the buffering of the data filtering PCs
becomes apparent: the unfiltered data set on disk includes data
that were recorded before the GRB was detected by the satellite,
and could include an early neutrino signal.

\section{GRB data analyses} 

The GRB data analysis can be done in two alternative ways.
The standard way is based on real-time filtered data and
reconstruction of the muon trajectory. 
The reconstruction is based on a five parameter fit, including the two
direction angles~\cite{aart}.
 
\begin{figure}[t!]
\centering
\includegraphics[width=3.0in]{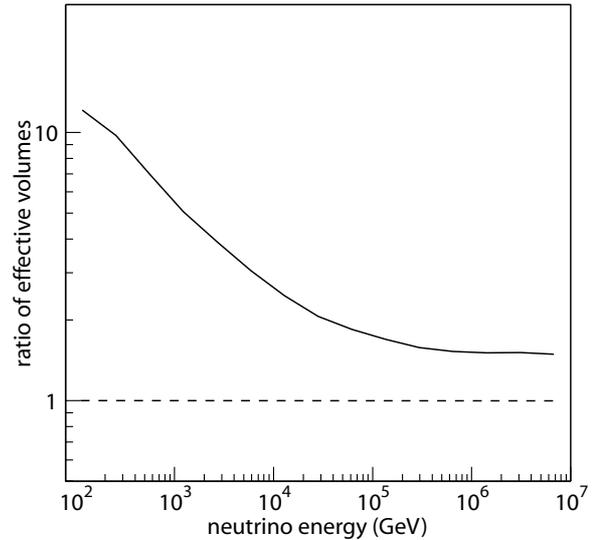}
\caption{The gain in detection efficiency for GRBs as a function of
neutrino energy (simulation). The solid line shows the result obtained
with the satellite triggered data taking, in combination with the
analysis method that uses the position of the GRB, relative to
the result obtained with the standard data taking and standard
analysis method (dashed line).}
\label{fig:fig04}
\end{figure}

The alternative method is based on the unfiltered data saved on disk
after a GRB alert. 
This analysis takes into account the position of the GRB on the sky, 
which is also provided by the GCN. 
Since these data do not need to be processed in real time, a much
lower detection threshold can be applied than is done for the standard data filtering. 
In the GRB analysis with unfiltered data, at least 6 time-position
correlated photons are required, compatible with a muon travelling in
the same direction as a neutrino from the GRB. 
In this way, the analysis method is only sensitive to a physics signal
from a specific GRB. 
The position of the GRB on the sky is also used to constrain the direction
angles in the fit. 
The same fit is repeated using many alternative directions, covering 
the opposite hemisphere and the downward hemisphere.
A cut on the likelihood ratio between 
the result of the first fit and 
the best result of all fits using the alternative directions  
is applied in order to select neutrinos coming from the GRB.
The whole analysis is thus reduced to a simple counting experiment.
In addition, the remaining background is low due to the short
duration of the GRB.  
The gain in detection efficiency over the standard method
is shown in Fig.~\ref{fig:fig04}.
This result is obtained using a simulation of the detector response to
muons, originating from neutrinos from a specific GRB.
The gain is expressed as the ratio of effective volumes 
(the volume in which a neutrino interaction produces a detectable
muon), and is shown relative to the result obtained with
the standard data taking and standard analysis method, using the same
simulated data.
The increased detection efficiency that is obtained with the unfiltered GRB data
is due to the lower detection threshold.
A higher threshold is required in the standard data taking method in order to
process the data in real time, which leads to an unavoidable detection inefficiency. 

For GRBs above the \antares\ horizon, for which a neutrino signal is
more difficult to find, an analysis method is applied to detect muons
induced by high-energy gamma rays from the source.  
The detection principle is presented in reference~\cite{goulven}.
A similar gain in detection efficiency can be expected when applying 
a specialised analysis to the unfiltered data saved on disk after a GCN
alert.

\section{Conclusions}  

The unique features of the \antares\ data acquisition system, 
in combination with the real-time distribution of GRB alerts
by the GCN, make it possible to trigger in real time on GRB alerts.
The \antares\ detector also has the possibility to buffer a large amount of data, 
resulting in very fast, and even negative, response times to GRB alerts. 
It is foreseen that the future, much larger, neutrino telescope KM3NeT~\cite{km3net}
will be designed such that it can trigger on GRB alerts in the same way.
These satellite triggered data lead to a significant increase in the sensitivity to neutrinos
from GRBs. 
Therefore the availability of networks like the GCN are very important
for neutrino telescopes. 
It is, however, imperative that the GRB alerts are distributed
within a few tens of seconds in order to maximise the discovery potential
for the detection of neutrinos from GRBs.


\begin{thebibliography}{99}

\bibitem{grbs}
E. Waxman, J. Bahcall, Phys. Rev. Lett {\bf 78} (1997) 2292;
P. M\'{e}sz\'{a}ros, E. Waxman, Phys. Rev. Lett. {\bf 87} (2001)
171102; 
C. Dermer, A. Atoyan, Phys. Rev. Lett. {\bf 91} (2003) 071102;
S. Razzaque, P. M\'{e}sz\'{a}ros, E. Waxman, Phys. Rev. Lett. {\bf 90}
(2003) 241103.

\bibitem{gcn}
http://gcn.gsfc.nasa.gov/

\bibitem{daq}
J. A. Aguilar {\textit{et al.}}, Nucl. Instrum. Meth. {\bf A570} (2007) 107.

\bibitem{aart}
A. Heijboer {\textit{et al.}}, Reconstruction of Atmospheric Neutrinos in \antares, these proceedings.

\bibitem{goulven}
G. Guillard {\textit{et al.}}, Gamma ray astronomy with \antares, these
proceedings. 

\bibitem{km3net}
U. F. Katz, Nucl. Instrum .Meth. {\bf A567} (2006) 457.

\end{thebibliography}
\end{document}